\documentclass[12pt]{article}

\usepackage{graphicx}
\usepackage{amssymb}
\usepackage{stmaryrd}

\usepackage{natbib}

\usepackage{my_definitions}

\usepackage{setspace}

\title{On the dependence of third- and fourth-order moments on stability
in the turbulent boundary layer}

\author{A. Maurizi\\
Institute of Atmospheric Sciences and Climate\\
via Gobetti 101, I-40129 Bologna, Italy\\
e-mail: \texttt{a.maurizi@isac.cnr.it}
}

\begin{document}
\maketitle
\begin{abstract}
In this short review it is suggested that the relationship between
third- and fourth-order moments of turbulence in the atmospheric
boundary layer depends on stability. This can explain some differences
among datasets, and provides a key point for modelling improvement.
\end{abstract}
\section{Introduction}

Data on the third- and fourth-order moments of turbulent velocities in
boundary layers have been collected for many years, not only in the
atmospheric boundary layer (ABL), where their evaluation is somewhat difficult
\citep{lenschow_etal-jaot-1994}, but also in laboratory experiments,
where the relationship between odd-order moments and their next
even-order was studied by \citet{durst_etal-tsf-1987}, for instance.

Some recent papers \citep{ferrero_etal-jas-2004, cheng_etal-jas-2005}
discuss the role of high order moments in parameterisations of the
ABL, often referred to as non-local models in that
space derivatives of high order moments are used to model lower orders.

The role of third order moments has long been recognised, at least in
the Convective Boundary Layer (CBL), as connected to non-local transport
properties \citep{wyngaard_etal-pf-1991}. As regards the fourth-order
moments, in the lack of any further information, the Millionchikov
hypothesis \citep[p.  241]{monin_yaglom-1971} is usually invoked.  This
means that for vanishing skewness that tends to zero, the normalised
fourth order moments (kurtosis) tend to their Gaussian values.

The question is whether the available data support this hypothesis, to
what extent and under which conditions.

\section{Review of data and parameterisations}

In the study of high-order moments, particular interest is devoted to
the normalised moments, \ie skewness ($S$) and kurtosis ($K$) and 
relationships between them.
\citet{maurizi_etal-ae-1999}, and subsequently
\citet{tampieri_etal-invento-2000}, collected and analysed data from
literature on horizontal and vertical velocity skewness and kurtosis,
considering a variety of turbulence generation mechanisms.
Furthermore, recent papers focus attention in particular on the CBL:
new measurements from aircraft \citep[GH hereafter]{gryanik_etal-jas-2002}
and with remote sensing acoustic techniques (SODAR)
\citep[AMT hereafter]{alberghi_etal-jam-2002} have been presented.

\begin{figure}
\begin{center}
\includegraphics[width=1.0\textwidth]{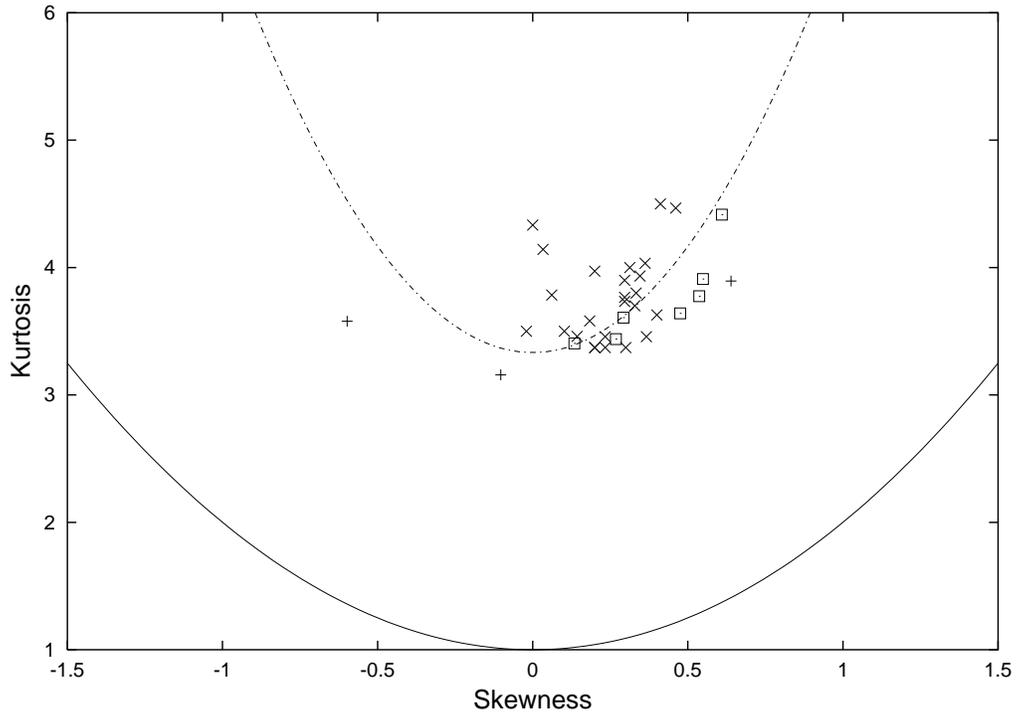}
\end{center}
\caption{Skewness and kurtosis for shear dominated boundary layers
\citep{tampieri_etal-invento-2000}. Continuous line represents the statistical
limit, while the dashed line is \Eqref{SK-tma} with $\alpha_0=3.3$.
(reprinted courtesy of \ldots)}
\label{SK-shear.fig}
\end{figure}

Different simple parameterisations were proposed for the $S$-$K$
relationship.  \citet{tampieri_etal-invento-2000}, first, and subsequently
\citet{maurizi_etal-ftc-2001}, proposed a parameterisation based on the
observation that a statistical limit exists in the $S$-$K$ space
\citep{kendall-stuart}, namely
\be
K \geq \subs{K}{lim}=S^2 +1\,.
\label{SK-limit}
\ee
This limit shapes the structure of the $S$-$K$. Thus, the pair $(S,\tilde
K)$ with $\tilde K=K(S^2+1)^{-1}$ can be taken as the natural
coordinate system for the $S$-$K$ space. The simplest model (zero
order) based on this observation can be built assuming a constant
$\tilde K$ and, therefore,
\be
K = \alpha_0 (S^2 +1)\,.
\label{SK-tma}
\ee
Fitting \Eqref{SK-tma} to data, \citet{tampieri_etal-invento-2000}
found, for the vertical velocity component, $\alpha_0=3.3$ for shear dominated
boundary layers (see \Figref{SK-shear.fig}) and $\alpha_0=2.5$ in the CBL
(AMT), adding
new data to the \citet{tampieri_etal-invento-2000} dataset, confirmed
the CBL result, giving $\alpha_0=2.4$.

GH
found that a mass-flux assumption for the
CBL exactly results in $K=\subs{K}{lim}$ (see \Eqref{SK-limit}) as a
relationship between $S$ and $K$. In fact, it is known that this
relationship only holds for two-value processes. They used, as a
generalisation, the form
\be
K=\alpha_0(\beta S^2+1)\,.
\label{SK-gh}
\ee
It is worth noting that \citet{lewis_etal-ncc-1997} proposed a
relationship of the same form for concentration data, based on the ideas
expressed by \citet{chatwin_etal-jfm-1990}. It can be observed
that, for $|S|<1$, \Eqref{SK-gh} is consistent with a second order model
in the $(S,\tilde K)$ space, namely
\be
\tilde K=\alpha_0+\alpha_1 S + \alpha_2
S^2\,,
\label{SK-second-order}
\ee
with $\alpha_1=0$ and $\alpha_2=\alpha_0(\beta-1)$.
Because the dataset of
GH dataset shows
strictly leptokurtic cases \citep[see also][]{cheng_etal-jas-2005},
they selected $\alpha_0=3$, $\alpha_1=0$ and $\alpha_2=-2$ in
\Eqref{SK-second-order}, and thus assumed Gaussianity for symmetric
distributions.

This assumption is not in agreement with the AMT
dataset, which shows the presence of a
great deal of data below $K=3$. However the data in
GH refer to a particular case (cold air
outbreak over the ocean) and fall within the area covered by the
AMT dataset, which collects measurements from
presumably different environmental conditions.

The question remains as to whether this subset is simply incomplete with
respect to a full convective behaviour, or whether it reflects the fact
that the production mechanism determines the $S$-$K$ relationship.
In fact, while the presented data are nominally taken in convective
conditions,
a concurrence of different production mechanisms (shear and
buoyancy) is actually expected. For instance, the Monin-Obukhov similarity
theory states that for $z \ll |L|$ shear dominates over buoyancy, and it
is presumable that data taken at different $z/L$ present a different
balance between the two mechanisms. Furthermore, the properties of
measurements taken across the CBL depth as a whole could depend on
$|L|/z_i$, which gives a measure of the fraction of CBL where buoyancy
cannot be considered as the only production mechanism.

The two values $\alpha_0=2.4$ to $\alpha_0=3.3$ in \Eqref{SK-tma} suggest
that the Gaussian case ($K\to3$ as $S\to0$) may occur for the vertical
turbulent velocity as a transition between the shear and convective
production mechanisms (for instance, in the CBL at $z < \subs{L}{MO}$).

This observation can be expressed by letting $\alpha_i$ be
a function of, at least, the Richardson number $\mbox{Ri}$, thus re-writing
\Eqref{SK-tma} as
\be
K=\alpha_0(\mbox{Ri})(S^2+1)
\label{k-general}
\ee
with the constraint $\alpha_0(\mbox{Ri}) > 1$.  It can be argued that
$\alpha_0$ increases as $z/L_{MO}$ increases from negative values to
zero, consistently with \citet[their Table 1]{anfossi_etal-blm-1997}.


\section{More on stability effects}
\begin{figure}
\begin{center}
\includegraphics[width=1.0\textwidth]{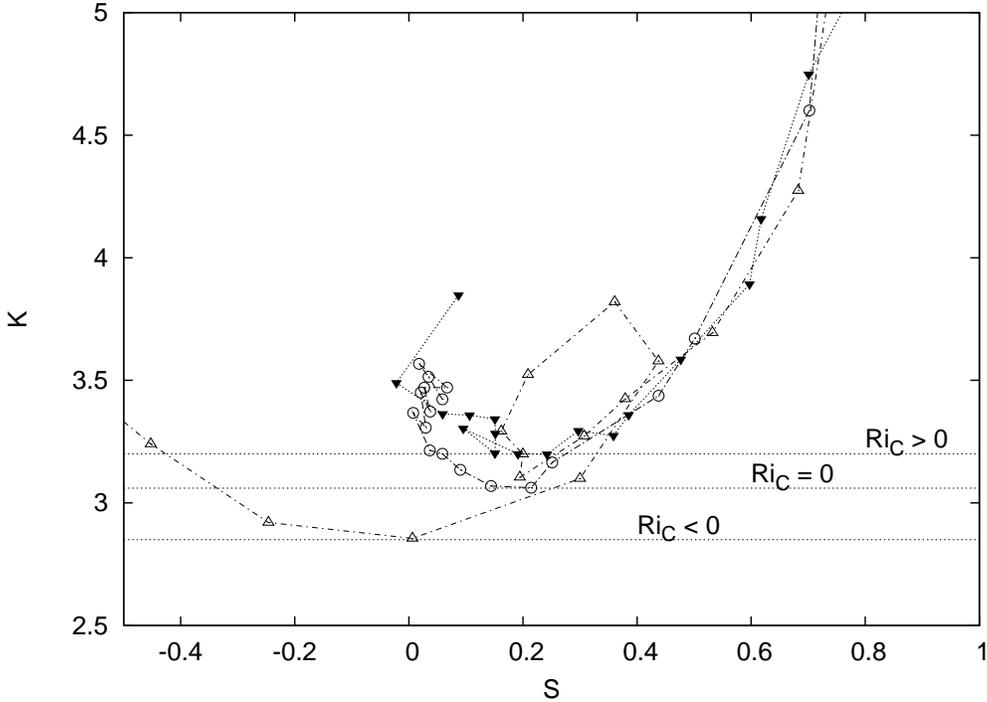}
\end{center}
\caption{Data of $S$ and $K$ for vertical velocity measured in turbulent
boundary layer over an obstacle at three different positions $x$ with respect
to the obstacle top. Full triangle: $x=0$ ($\subs{\mbox{Ri}}{C}>0$),
open circle: $x=L$ ($\subs{\mbox{Ri}}{C}=0$), open triangle: $x=-\infty$
($\subs{\mbox{Ri}}{C}<0$).}
\label{hill}
\end{figure}

\begin{table}
\begin{center}
\begin{tabular}{|p{0.15\textwidth}|p{0.15\textwidth}|p{0.15\textwidth}|}
\cline{1-3}
$R_c<0$&$R_c=0$&$R_c>0$\\
\cline{1-3}
2.8&3.0&3.2\\
\cline{1-3}
\end{tabular}
\end{center}
\caption{Minimum value of $K$ measured in three different stability
conditions.}
\label{table}
\end{table}

It would be interesting to investigate this issue further if certain
details on the measurements were available. In order to overcome the
unavailability of such information, we consider a flow over a simple
obstacle, and use the analogy between buoyancy and streamline curvature
\citep[see, \eg][]{bradshaw-jfm-1969,baskaran-jfm-1991} to investigate
on the role of stability in determining the character of the $S$-$K$
relationship.

In a neutral turbulent flow with streamline curvature, such as a flow
over a hill, it is possible to define a curvature Richardson
number $\subs{\mbox{Ri}}{C}$ \citep{wyngaard-phd-1967} as
\be
\subs{\mbox{Ri}}{C}=\frac{\displaystyle\frac{2U}{R}}{\displaystyle\frac{\partial
U}{\partial z}+\displaystyle\frac{U}{R}}
\ee
where $U$ is the mean velocity module and $R$ is the streamline
curvature radius. Positive $\subs{\mbox{Ri}}{C}$ corresponds to a
dumping term in the turbulent energy budget (\eg on the hill top)
expressed in streamline coordinates, while negative $\subs{\mbox{Ri}}{C}$
represents an unstable contribution, \eg near the hill base.

An experiment carried out in the Enflo ``A'' wind tunnel
focused attention on third- and fourth-order
turbulence statistics
. A turbulent boundary layer was
generated upstream of a sinusoidal, two-dimensional hill with aspect
ratio $H/L=5$, where $H$ is the obstacle height and $L$ is half the
total width. Turbulence was measured with a hot-wire anemometer at 7
different positions streamwise, at 20 vertical levels. Measurements were
recorded for times long enough to produce reliable statistics up to the
fourth order.

\Figref{hill} shows data of $S$ and $K$ for three profiles measured far
upstream, on the hill top and on the lee side at the hill foot. Those
regions correspond to $\subs{\mbox{Ri}}{C}=0$, $\subs{\mbox{Ri}}{C}>0$
and $\subs{\mbox{Ri}}{C}<0$, respectively. Apart from the specific
structure revealed by the measurements in the unstable case for $S$ in
the range $0.2$--$0.4$, the form of the $S$-$K$ relationship is similar
for the three cases and exhibits a minimum value.

We use this minimum as a rough indicator of the different features
of the $S$-$K$ relationship. This seems a reasonable parameter in that
it could discriminate data reported by
GH from other data in the
AMT dataset.

The minimum measured values of $K$ are reported in \Tabref{table} for
three different cases of $\subs{\mbox{Ri}}{C}$ estimated
from a similar flow \citep{maurizi_etal-ncc-1997}.

Although measured profiles are largely influenced by the specific
dynamics of flow over the obstacle and, in particular, by non local
equilibrium in the lee side, results are in qualitative agreement with
the proposed dependence of expansion coefficients on stability
(\Eqref{k-general}).

\section{Conclusions}

The $S$-$K$ relationship is modelled by an expansion in an appropriate space
and some parameterisations in literature have been reduced to this scheme.

In this frame, comparing data for shear- and convective-dominated
boundary layers, it is found that model constants should depend on
stability.
In particular, the Gaussian case can occur in intermediate situations
between the purely shear- and purely convective-dominated boundary layer.

Furthermore, an examination of the differences between two datasets for the
convective case suggests that that there could be a variety of
behaviours for different balances between shear and buoyancy production.

As an example, some data measured in a wind tunnel experiment have shown
the possibility that the suggested dependence can be confirmed. It is
worth pointing out that this dataset if far from exhaustive with respect
to the problem studied and it has been used merely as a indicator.

Further measurements of high-order moments of wind velocity in really
different stability conditions are required in order to provide a more
quantitative response to the problem.

\section*{Acknowledgements}

The author would like to thank Francesco Tampieri for invaluable helpful
discussions. The support of the ``Italia-USA Project on Climatic
Change''  is also acknowledged.

\bibliographystyle{jam}
\bibliography{abbr,all}

\end{document}